\documentstyle[12pt]{article}
\let\LARGE=\Large
\let\Large=\large
\let\large=\normalsize
\newcommand{\be}[3]{\begin{equation}  \label{#1#2#3}}     % non-hyper
\newcommand{\ee}{ \end{equation}}
\newcommand{\ba}{\begin{array}}
\newcommand{\ea}{\end{array}}

\def\beq{\begin{equation}}
\def\eeq{\end{equation}}
\def\beqa{\begin{eqnarray}}
\def\eeqa{\end{eqnarray}}

\begin{document}

\begin{titlepage}
\begin{center}
\hfill THU-98/16\\
\hfill {\tt hep-th/9804064}\\

\vskip 3cm

{ \LARGE \bf Charged Heterotic Black-Holes

in Four and Two Dimensions}

\vskip .3in

{\bf Gabriel Lopes Cardoso
}\footnote{\mbox{Email: \tt 
cardoso@fys.ruu.nl}}
\\
%\vskip 1.2cm
\vskip 1cm

{\em Institute for Theoretical Physics, Utrecht University,
3508 TA Utrecht, Netherlands}\\

\vskip .1in

\end{center}

\vskip .2in

\begin{center} {\bf ABSTRACT } \end{center}

We consider four-dimensional charged black-holes occuring
in toroidally compactified heterotic string theory, whose ten-dimensional
interpretation involves a Kaluza-Klein monopole and a five-brane.
We show that these four-dimensional black-holes can be connected
to two-dimensional charged heterotic black-holes upon removal
of the constants appearing in the harmonic functions associated with
the Kaluza-Klein monopole and the five-brane.

\vskip 6cm
April 1998\\
\end{titlepage}
\vfill
\eject

\newpage

It was pointed out by Hyun 
in \cite{hyun} that five-dimensional 
black-holes occuring in toroidally compactified type II string theory 
are related 
to two-dimensional black-holes
by U-duality.  In \cite{teo} it was shown that the
associated entropies agree.  Here, we shall consider
a particular class of extremal four-dimensional black-holes occuring 
in toroidally
compactified heterotic string theory and we will
connect it to a class of extremal two-dimensional
heterotic black-holes.  The two-dimensional black-holes
in question \cite{guinayo,gipe} are derived from the heterotic string
target-space action 
\beqa
{1\over16\pi G_2}\int d^2x\,\sqrt{-g}\,
{\rm e}^{-2\phi}\left\{R+4(\nabla\phi)^2- \frac{1}{4}F^2
+c\right\} \;\;\;.
\label{twoact}
\eeqa
We will show that the associated entropies match. It should be 
straightforward to extend this connection to the case of non-extremal 
black-holes.

The extremal four-dimensional heterotic black-holes we will consider 
are the
ones obtained by compactifying the heterotic
ten-dimensional solution describing
the intersection of a string, a five-brane, a Kaluza-Klein monopole
and a wave \cite{cvetsey}.  The ten-dimensional line element is given by
\beqa
ds^2_{10} &=&  \frac{1}{H_1} \left( - dt^2 + dy^2 + (H_0-1) (dt + dy)^2\right)
 + 
H_5 H_{KK} dx_i^2 + \frac{H_5}{H_{KK}} (dz + A_i dx^i)^2 \nonumber\\
&+& 
dx^2_6 + dx^2_7 + dx^2_8 + dx^2_9 \;\;\;,\;\;\; r^2 = x_i^2\;\;\;,\;\;\;
i = 1,2,3
\label{lineten}
\eeqa
where the $H_i = 1 + \frac{Q_i}{r}$ ($i = 0,1,5,KK$) denote the
harmonic functions associated with the wave, the string, the five-brane and
the Kaluza-Klein monopole, respectively.  Also
$F_{ij} = \partial_i A_j - \partial_j A_i = \varepsilon_{ijk} 
\partial_k H_{KK}$. 
The ten-dimensional dilaton and the 
antisymmetric tensor are given by
\beqa
e^{-2 \phi} = \frac{H_1}{H_5} \;\;\;,\;\;\; B_{ty} = \frac{1}{H_1}
\;\;\;,\;\;\; H_{z ij} = \frac{1}{2} \varepsilon_{ijk} \partial_k H_5 \;\;\;.
\label {dilaten}
\eeqa
Dimensionally reducing in the $y, z, x_6, x_7, x_8 , x_9$ directions
and going to the Einstein frame yields the four-dimensional metric
\beqa
ds^2_4 = - \frac{1}{\sqrt{H_0 H_1 H_5 H_{KK}}} \, dt^2 + 
\sqrt{H_0 H_1 H_5 H_{KK}} \, dx^2_i \;\;\;.
\eeqa
This describes an extremal four-dimensional black-hole, and its entropy is
given by \cite{cveyou}
\beqa
S = \frac{A}{4G_4} = \frac{\pi}{G_4} \sqrt{Q_0 Q_1 Q_5 Q_{KK}} \;\;\;,
\label{entrofour}
\eeqa
where $A$ denotes the area of the horizon and $G_4$ denotes
Newton's constant in four dimensions.

Before relating this four-dimensional black-hole to an extremal 
two-dimensional black-hole, let us first link the near-horizon geometry
of the former 
to the geometry of a three-dimensional BTZ black-hole, 
as in \cite{hyun,sfeske,balalars}.  This is achieved by 
dimensionally reducing (\ref{lineten}) in the 
$z, x_6, x_7, x_8 , x_9$ directions and by 
going to the Einstein frame. In the near-horizon region ($Q_i/r \gg 1$) 
the resulting five-dimensional metric is given by
\beqa
ds^2_5 = \frac{r}{(Q_1Q_5Q_{KK})^{\frac{1}{3}}} \left( - dt^2
+ dy^2 + \frac{Q_0}{r} (dt + dy)^2 \right) + (Q_1Q_5Q_{KK})^{\frac{2}{3}}
( \frac{dr^2}{r^2} + d\Omega_2^2 ) \;\;\;.
\eeqa
The five-dimensional dilaton and the
$zz$ component of the metric are approximately constant and given by
$e^{-2 \phi} = Q_1 / \sqrt{Q_5 Q_{KK}}$ and
$g_{zz}=Q_5/Q_{KK}$, respectively.  The five-dimensional metric   
can be rewritten as \cite{balalars}
\beqa
ds^2_5 &=& ds^2_3 + \frac{l^2}{4} d \Omega^2_2 \;\;\;,\;\;\; 
l = 2 (Q_1 Q_5 Q_{KK})^{\frac{1}{3}} \;\;\;, \nonumber\\
ds^2_3 &=& - N^2 d \tau^2 + \frac{d \rho^2}{N^2} + \rho^2 (d \phi 
+ N_{\phi} d \tau)^2 \;\;\;,
\eeqa
where
\beqa
N^2 &=& \frac{\rho^2}{l^2} - M + \left(\frac{8 G_3 J}{2 \rho} \right)^2 
\;\;\;,\;\;\; N_{\phi} =  \frac{8 G_3 J}{2 \rho^2} \;\;\;,\;\;\;
M = \frac{8 G_3 J}{l} = \frac{4 Q_0}{l^3} \;\;\;,
\eeqa
and where $\tau = l t,\, \rho^2 = 2 ( r + Q_0)/l, \,
\phi = y ,\, 0 \leq \phi \leq 2 \pi$.  
This geometry
describes a space which is the product of a two-sphere of radius $l/2$ and
an extremal BTZ black-hole with mass $M$ and angular momentum $J$.  
Note that the BTZ part and the sphere are completely decoupled.
Inspection of (\ref{lineten}) and (\ref{dilaten})
shows that the two-sphere supports a $U(1)$ gauge
field.  The extremal BTZ black-hole carries 
entropy \cite{banados} 
\beqa
S = \frac{\pi}{4G_3} \sqrt{2 l^2 M} \;\;\;,
\label{entrothree}
\eeqa
where $G_3$ denotes the three-dimensional Newton's constant.
Using that the three and four-dimensional Newton's constants
are related by \cite{balalars} $\frac{1}{G_3} = \frac{1}{G_4}
\frac{l^2}{2}$
it follows that the BTZ entropy (\ref{entrothree}) agrees with the
macroscopic entropy (\ref{entrofour}) of the four-dimensional black-hole.

Next consider the five-dimensional metric in the string frame, obtained
by reducing (\ref{lineten}) over
$z, x_6, x_7, x_8 , x_9$:
\beqa
ds^2_5 =  \frac{1}{H_1} \left( - dt^2 + dy^2 + (H_0-1) (dt + dy)^2\right)
 + 
H_5 H_{KK} dx_i^2  \;\;\;,\;\;\; g_{zz} = \frac{H_5}{H_{KK}} \;\;\;.
\label{linefive}
\eeqa
The five-dimensional dilaton reads 
$e^{-2 \phi} = H_1 / \sqrt{H_5 H_{KK}}$.  Let us now drop the constant
part in the harmonic functions $H_5$ and $H_{KK}$ associated with the 
five-brane and the Kaluza-Klein monopole, so that the 
resulting space-time is no longer asymptotically flat.
Below we will argue that the
constant part in these harmonic functions should be removable by a sequence
of $S$ and $T$-$S$-$T$ duality transformations in four dimensions.
Let us furthermore identify $H_0 = H_1$.  Then, the line element
(\ref{linefive}) turns into
\beqa
ds^2_5 =  \frac{1}{H_1} \left( - dt^2 + dy^2 + (H_1-1) (dt + dy)^2\right)
 + 
Q_5 Q_{KK} (\frac{dr^2}{r^2} + d \Omega_2^2)  \;\;\;.
\label{nlinefive}
\eeqa
Note that 
$g_{zz}= Q_5/Q_{KK}$ is now constant.
Dimensionally reducing (\ref{nlinefive}) in the $y$ direction yields
\beqa
ds^2_4 = - \frac{1}{H_1^2} dt^2 + Q_5 Q_{KK} \frac{dr^2}{r^2} 
+  Q_5 Q_{KK} \, d \Omega_2^2
\;\;\;,\;\;\; e^{- 2 \phi} = \frac{r + Q_1}{\sqrt{Q_5 Q_{KK}}} \;\;\;.\;\;\;
g_{yy} =1 \;\;\;.
\label{linett}
\eeqa
This metric describes the product of two two-dimensional
spaces, one of
which we will relate to a two-dimensional black-hole 
and one which is a two-sphere with radius $\sqrt{Q_5 Q_{KK}}$.  
Inspection of (\ref{lineten}) and (\ref{dilaten}) shows that each of the
two-dimensional geometries 
supports a $U(1)$ gauge field.  Both two-dimensional
parts are completely decoupled.  Note that since the internal components
of the metric $g_{yy}$ and $g_{zz}$ are constant, the only
non-trivial scalar field is the dilaton.

The solution (\ref{linett}) can now be related to the two-dimensional
heterotic black-hole solution of McGuigan, Nappi and Yost
\cite{guinayo} as follows.  The latter is described by
\beqa
ds^2_2 &=& - ( 1 - 2m e^{-Qx} + q^2 e^{-2Qx}) dt^2 + 
( 1 - 2m e^{-Qx} + q^2 e^{-2Qx})^{-1} dx^2 \;\;\;, \nonumber\\
e^{-2(\phi - \phi_0)} &=& e^{Qx} \;\;\;,\;\;\;
F_{tx} = \sqrt{2} Q q e^{-Qx} \;\;\;.
\label{bh2}
\eeqa
Here $m$ and $q$ are constants related to the mass 
and to the electric charge of the solution, with $m > 0$ and
$m^2 \geq q^2$.  $Q$ is a positive constant which determines
the central charge deficit $c= Q^2$ appearing in the action (\ref{twoact}).
The asymptotic flat region corresponds to $x = \infty$,
whereas the curvature singularity is at $x = - \infty$.  By changing the
spatial variable to $y = e^{-Qx}$, the solution (\ref{bh2}) becomes
\beqa
ds^2_2 &=& - q^2 (y-y_1)(y-y_2) \, dt^2 + \frac{1}{q^2 Q^2} \frac{dy^2}
{y^2 (y - y_1) (y-y_2)} \;\;\;, \nonumber\\
e^{-2(\phi - \phi_0)} &=& \frac{1}{y} \;\;\;,\;\;\;
A_{t} = \sqrt{2} q y \;\;\;.
\label{twobh}
\eeqa
The asymptotic flat region is now at $y=0$, the curvature singularity
is at $y = \infty$, while th two horizons are at $y_{1,2} = q^{-2}
( m \pm \sqrt{m^2 - q^2})$.  In the following, we will consider the extremal
case, which corresponds to $m =q$ and hence to $y_1 = y_2 = q^{-1}$.
By setting
\beqa
y = y_1 \; \frac{a^2}{r + a^2} = \frac{1}{q} \; \frac{a^2}{r + a^2}
\eeqa
the solution (\ref{twobh}) can be turned into
\beqa
ds^2_2 &=& - \frac{dt^2}{(1 + \frac{a^2}{r})^2} + \frac{1}{Q^2} 
\frac{dr^2}{r^2}
\;\;\;, \nonumber\\
e^{-2(\phi - \phi_0)} &=& \frac{q}{a^2} ( r + a^2)  \;\;\;,\;\;\;
A_{t} = \sqrt{2} \frac{a^2}{r+a^2} \;\;\;.
\label{nbh}
\eeqa
The two-dimensional $(t,r)$ part of (\ref{linett}) thus agrees with the charged
black-hole solution (\ref{nbh}) provided that the parameters $a,Q$ and $q$
appearing in (\ref{nbh}) are related to the charges $Q_1,Q_5$ and $Q_{KK}$
as follows: 
\beqa
a^2 = Q_1 \;\;\;,\;\;\; Q^2 = \frac{1}{Q_5 Q_{KK}} \;\;\;,\;\;\;
q e^{-2 \phi_0} = \frac{Q_1}{\sqrt{Q_5 Q_{KK}}} \;\;\;.
\label{param}
\eeqa
The field strength of the 
gauge field in (\ref{nbh}) agrees with $H_{rty} = \partial_r B_{ty}$ given in 
(\ref{dilaten}), up to a constant.

The entropy of the charged two-dimensional black-hole (\ref{twobh})
was computed in \cite{guinayo,napasq} using thermodynamic methods.
For an extremal black-hole, it is given by
\beqa
S = \frac{1}{4  G_2} e^{-2 \phi_0} m \;\;\;.
\label{entrotwo}
\eeqa
Here, $G_2$ denotes the two-dimensional Newton's constant, which was set
to $G_2 = 1/(16 \pi)$ in \cite{guinayo,napasq}.  Inspection of (\ref{linett})
shows that $G_2$ is related to 
Newton's constant $G_4$ in four dimensions by
\beqa
\frac{1}{G_2} = \frac{V}{G_4} \;\;\;,\;\;\; V = 4 \pi Q_5 Q_{KK} \;\;\;,
\label{g2g4}
\eeqa
where $V$ denotes the area of the two-sphere with radius $\sqrt{Q_5 Q_{KK}}$.
Inserting (\ref{g2g4}) as well as (\ref{param}) into the
entropy formula (\ref{entrotwo}) yields 
\beqa
S = \frac{1}{4  G_2} e^{-2 \phi_0} m = \frac{\pi}{G_4} Q_1 \sqrt{Q_5 Q_{KK}}
\;\;\;,
\eeqa
in agreement with the macroscopic formula (\ref{entrofour}) of the 
four-dimensional black-hole.

We note here that we expect each of the decoupled two-dimensional geometries
of (\ref{linett}) to be described by an exact CFT.  We refer to
\cite{steif,horne,john,sfettsey} for a discussion of the CFT description
of the
two-dimensional charged black-hole (\ref{bh2}).

We would now like to argue that there is a sequence of $S$ and 
$T$-$S$-$T$ duality transformations which can be used 
to shift the constant part of the harmonic functions $H_5$ and
$H_{KK}$.  Consider dimensionally reducing 
the ten-dimensional line element
(\ref{lineten}) in the $t, y, x_6, x_7, x_8, x_9$ directions, that is
consider compactifying it on a 
six-torus with Lorentzian signature.  The resulting four-dimensional metric 
in the Einstein frame and the four dimensional dilaton read
\beqa
ds^2_4 = 
H_{KK} dx_i^2 + \frac{1}{H_{KK}} (dz + A_i dx^i)^2 \;\;\;,\;\;\;
e^{-2\phi} = \frac{1}{H_5} \;\;\;.
\label{lineeu}
\eeqa
The antisymmetric 
tensor field $H_{zij}$ given in (\ref{dilaten}) can be dualised
to an axion field $b$, which can then be combined with $
e^{-2\phi}$ into $S_{\pm} = b \pm e^{-2\phi}$ \cite{bakas}.  
$S_{\pm}$ can undergo $SL(2,R)$ transformations.  The form of the line element
(\ref{lineeu}) makes it clear that, in the
case that $H_1=H_{KK}=1$, these $SL(2,R)$ transformations can be used
to shift away the constant part of $H_5$.  Similarly, 
in the case that $H_1 = H_5 =1$, 
there exists \cite{bakas}
a sequence of $T$-$S$-$T$ duality transformations which can be used to 
remove the constant term in $H_{KK}$.  In view of that, we expect that also
in the case where $H_1, H_5$ and $H_{KK}$ are all non-constant, there is
a certain 
sequence of $S$ and $T$-$S$-$T$ duality transformations which is capable
of removing
the constant terms in $H_5$ and $H_{KK}$.  If so, the solutions (\ref{lineten})
with and without constant terms in $H_5$ and $H_{KK}$ would be dual to one
another.

{\large \bf Acknowledgements}

\smallskip
\noindent
This work is supported by 
the European 
Commission TMR 
programme ERBFMRX-CT96-0045.


\begin{thebibliography}{111}

\bibitem{hyun} S. Hyun, {\it
U-duality between Three and Higher Dimensional Black-Holes}, 
{\tt hep-th/9704005}.

\bibitem{teo} E. Teo, {\it Statistical Entropy of Charged
Two-dimensional Black-Holes}, {\tt hep-th/9803064}.


\bibitem{guinayo} M. D. McGuigan, C. R. Nappi and S. A. Yost, 
{\it Nucl. Phys.} {\bf B375} (1992) 421, {\tt hep-th/9111038}. 

\bibitem{gipe} G. W. Gibbons and M. J. Perry, {\it Int. J. Mod. Phys.}
{\bf D1} (1992) 335, {\tt hep-th/9204090}.


\bibitem{cvetsey} M. Cvetic and A. Tseytlin, 
{\it Phys.. Lett.} {\bf B366} (1996) 95,
{\tt hep-th/9510097}.

\bibitem{cveyou} M. Cvetic and  D. Youm, {\it Phys. Rev.} {\bf  D53}
 (1996) 584, {\tt hep-th/9507090}.


\bibitem{sfeske} K. Sfetsos and K. Skenderis, 
{\it Microscopic Derivation of the Bekenstein-Hawking Entropy Formula for
Non-Extremal Black Holes}, {\tt hep-th/9711138}.

\bibitem{balalars} V. Balasubramanian and F. Larsen, 
{\it Near Horizon Geometry and Black-Holes in Four Dimensions}, 
{\tt hep-th/9802198}.

\bibitem{banados} M. Ba\~nados, C. Teitelboim and J. Zanelli,
{\it Phys. Rev. Lett.} {\bf 69} (1992) 1849, {\tt hep-th/9204099}.

\bibitem{napasq} C. R. Nappi and A. Pasquinucci, 
{\it Mod. Phys. Lett.} {\bf A7} (1992) 3337,
{\tt gr-qc/9208002}.

\bibitem{steif} N. Ishibashi, M. Li and A. R. Steif, {\it
Phys. Rev. Lett.} {\bf 67} (1991) 3336.


\bibitem{horne} J. H. Horne and G. T. Horowitz, {\it Nucl. Phys.} {\bf B368}
(1992) 444, {\tt hep-th/9108001}.



\bibitem{john} C. V. Johnson, {\it Phys. Rev.} {\bf D50} (1994) 4032,
{\tt hep-th/9403192}.

\bibitem{sfettsey} K. Sfetsos and A. A. Tseytlin,
{\it Nucl. Phys.}  {\bf B427} (1994) 245, {\tt hep-th/9404063}.




\bibitem{bakas} I. Bakas, {\it Phys. Lett.} {\bf B343} (1995) 103, 
{\tt hep-th/9410104}.





\end{thebibliography}
\end{document}